\documentclass[preprint,12pt,review]{elsarticle}

\usepackage{amssymb}
\usepackage{amsmath}
\newcommand{\bi}{\bibitem}

\journal{Physics Letters B}

\begin{document}

\begin{frontmatter}




\title{Exclusive production of pseudoscalar mesons  in neutrino--photon interactions}


\author{I. Alikhanov\corref{cor1}}

\cortext[cor1]{{\it Email address: {\tt ialspbu@gmail.com}},
\,{\it Telephone:} +78663875221, {\it Fax:} +78663875103}

\address{Institute for Nuclear Research of the Russian Academy of Sciences,
60-th October Anniversary pr. 7a, Moscow 117312, Russia}

\begin{abstract}
Exclusive production of the $\pi$ mesons in neutrino--photon interactions at low momentum transfer is studied within the standard model. The corresponding cross sections are calculated analytically. Potential astrophysical implications and the significance for testing the standard model are discussed. The presented formalism applies to other pseudoscalar mesons as well.

\end{abstract}


\end{frontmatter}


\section{Introduction}
Neutrino--photon  interactions can play an important role in astrophysical and cosmological phenomena such as stellar evolution, production of high energy cosmic rays, detecting the relic neutrino background. In the past few decades this type of interactions has attracted some interest and a definite progress  has been reached in this field \cite{chiu1960,cung1975,dicus1993,dicus1997,harris,seckel,Shaisultanov,dicus1999,abada1, abbas, abada2,Ioannisian,Aghababaie,
masso,Goyal,Chistyakov,jetp,Haghighat,newest,mine1,mine2}.

For example, it has been realized that the inelastic process $\nu\gamma\rightarrow\nu\gamma\gamma$ significantly dominates over elastic scattering $\nu\gamma\rightarrow\nu\gamma$ \cite{dicus1993,dicus1999,abada1,abada2}. In its turn, when the energy threshold of the electron--positron pair production is crossed, the reaction $\nu\gamma\rightarrow\nu e^+e^-$ becomes the dominant one~\cite{masso}.

Neutrino--photon reactions with relatively low energy thresholds as those mentioned above are of special interest for astrophysics. They are crucial for understanding processes of the energy loss by stars \cite{itoh}, especially the collapsing ones, when they may loose a significant part of their masses by the neutrino emission \cite{supernova}.

In the conservative estimation, temperature in the interiors of a supernova ranges from 10 MeV up to 100 MeV and $\nu\gamma$ reactions are therefore able to produce final states with masses as large as the mass of the $\pi$ meson.
Thus, in addition to the traditionally considered energy loss resulting from production of structureless particles in inelastic $\nu\gamma$ scattering~\cite{masso}, pair, photo-, plasma, bremsstrahlung and recombination neutrino processes~\cite{itoh}, one should also take into account the possibility of excitation of states like the $\pi$ mesons by neutrinos propagating through the star.

Due to above reasons the exclusive production of the $\pi$ mesons in the reactions

\begin{eqnarray}
\nu_e\gamma\rightarrow e^-\pi^+,\label{reac1}
\\
\nu_l\gamma\rightarrow \nu_l\pi^0 \label{reac2}
\end{eqnarray}

is studied  in the present paper. In (\ref{reac2}) $l$ stands for $e$, $\mu$ and $\tau$. 

It should be noted that the reactions (\ref{reac1}) and (\ref{reac2}) have not only astrophysical implications but also provide crucial tests of the standard model. Actually, (\ref{reac1}) is the crossed reaction of the radiative pion decay $\pi ^{+} \to e^{+}\nu_{e}\gamma$ ($\pi_{e2\gamma}$). The latter and other light pseudoscalar decays are extensively discussed in the literature as possible tests of the $V-A$ structure of the weak interaction (see, for example, \cite{Chen2011} and the references therein).
Processes closely related to (\ref{reac2}) are also candidate solutions of the problem of the excess of electron-like events observed by the MiniBooNE Collaboration in a search for $\bar \nu_\mu \rightarrow \bar \nu_e$ oscillations which arouses a hot discussion today \cite{Harvey2007,miniboone}.

\section{Pion production and its crossed reactions \label{sec:charged}}

\subsection{Exclusive production of a charged pion in $\nu_e\gamma$ interactions\label{subsec:charged}}

\begin{figure}
\centering
\resizebox{1.2\textwidth}{!}{%
\includegraphics{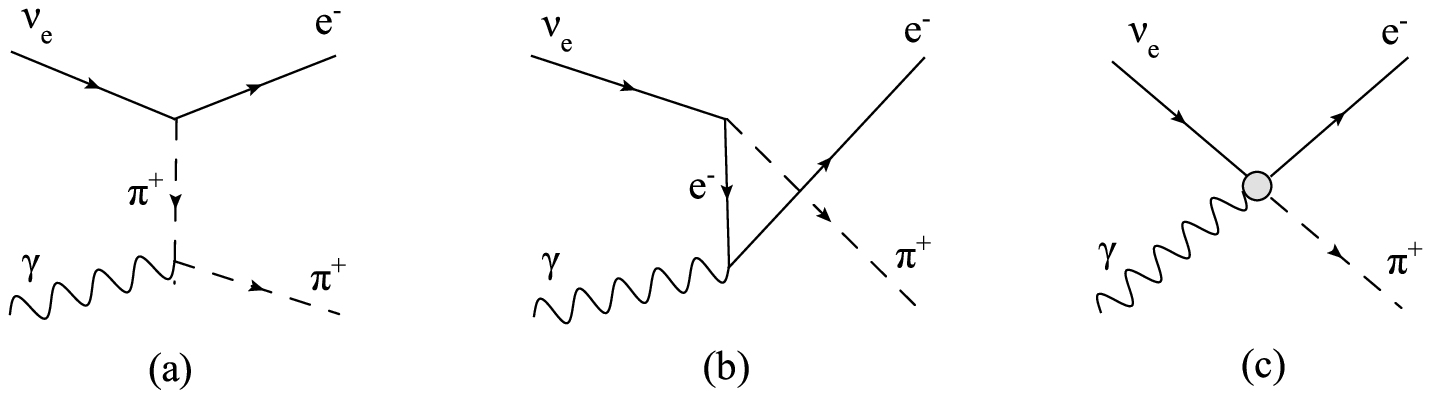}
}\caption{}
\label{fig1}
\end{figure}

Let us consider the reaction (\ref{reac1}) at low momentum transfer satisfying the condition~$q^2\ll~M^2_W$ so that one may use the Fermi approximation.  The corresponding Feynman diagrams in the leading order are presented in Fig.~\ref{fig1}.

One can show that the contributions of the diagrams (a) and (b) are helicity suppressed being proportional to the electron mass $m_e$ exactly as in the case of the $\pi_{e2}$ decay \cite{Chen2011}. For this reason we will neglect them in the subsequent calculations keeping only the diagram (c) which is free of the suppression and depends on the vector and axial-vector weak hadronic currents characterizing the structure of the pion \cite{Chen2011,Bryman}. 
Then, noting that (\ref{reac1}) is a crossed reaction of the decay $\pi_{e2\gamma}$ one can find the amplitude just by crossing the corresponding result of \cite{Chen2011}: 

\begin{equation}
M=-i{\frac{G_{F}}{\sqrt{2}}}V_{ud}\varepsilon_{\mu }
\bar{u}(p_{e})\gamma _{\alpha }(1-\gamma
_{5})u(p_{\nu})
\left[e\frac{F_{A}}{m_{\pi}}(-g^{\mu \alpha }p_{\pi}\cdot
q+p_{\pi}^{\mu }q^{\alpha })+ie\frac{F_{V}}{m_{\pi}}\epsilon ^{\mu \alpha \beta
\lambda }q_{\beta }p_{\pi \lambda }\right], \label{ampl_decay}
\end{equation}   

where $V_{ud}$ is the Cabibbo--Kobayashi--Maskawa (CKM) matrix element, $\varepsilon_{\mu}$ is the photon polarization vector,
$p_{\pi}$, $p_{e}$, $p_{\nu}$, and $q$
are the four momenta of $\pi^{+}$, $e^{-}$,
$\nu_{e}$ and $\gamma$, respectively, $F_{A}$ and $F_{V}$ are
the axial-vector and vector form factors.

Squaring (\ref{ampl_decay}), averaging over the two spin states of the initial photon, summing over the final state spins gives

\begin{equation}
|M|^2=-\frac{e^2G^2_{F}}{2m^2_{\pi}}|V_{ud}|^2\,t\left(u^2|F_V+F_A|^2+s^2|F_V-F_A|^2\right), \label{ampl_d_squared}
\end{equation}   
where the conventional Mandelstam variables are used: $s=(p_{\nu}+q)^2$, $t=(p_{\nu}-p_{e})^2$, $u=(p_{\nu}-p_{\pi})^2$.

Knowing (\ref{ampl_d_squared}) one obtains the differential cross section

\begin{equation}
\frac{d\sigma}{dt}=-\frac{\alpha G^2_{F}}{8m^2_{\pi}}|V_{ud}|^2\,t\left(\frac{u^2}{s^2}|F_V+F_A|^2+|F_V-F_A|^2\right), \label{cross_sect}
\end{equation}

where $\alpha$ is the fine structure constant.

Integration of (\ref{cross_sect}) over the physically possible values of $t$ ($m^2_{\pi}-s\leq t\leq 0$) yields the total cross section of the reaction (\ref{reac1}):  

\begin{equation}
\sigma=\frac{\alpha G^2_{F}}{96m^2_{\pi}}|V_{ud}|^2\,s^2\left(1-\frac{m^2_{\pi}}{s}\right)^2\left(\left(1-\frac{m^2_{\pi}}{s}\right)^2|F_V+F_A|^2+6|F_V-F_A|^2\right). \label{cross_sect_tot}
\end{equation}

Note that though the formfactors depend, in general, on $t$, in the present analysis they are taken to be constants since we deal with reactions proceeding at conditions comparable to the case of the decay $\pi_{e2\gamma}$ ($t\sim m_{\pi}^2$) \cite{Chen2011}. Thus, throughout this paper we set $F_V=0.0272$ and $F_A=0.0112$ \cite{Chen2011}.

\begin{figure}
\centering
\resizebox{0.9\textwidth}{!}{%
\includegraphics{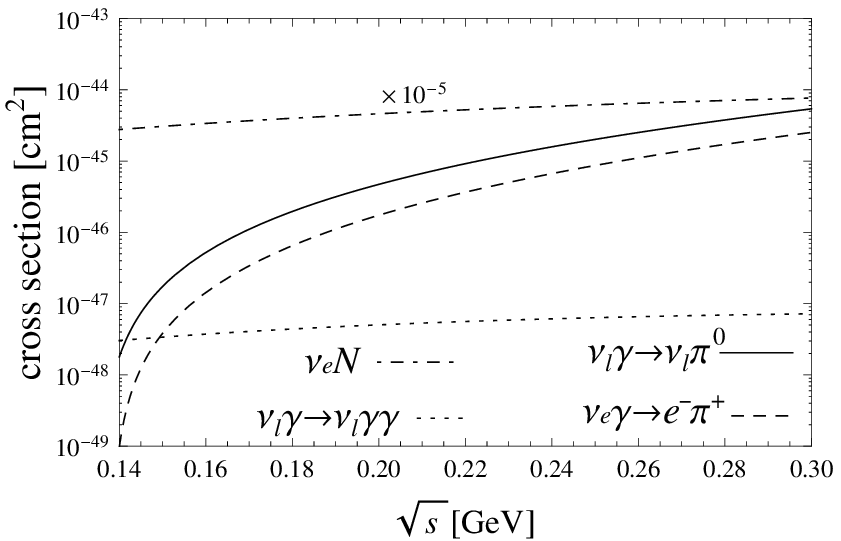}
}\caption{}
\label{fig2}
\end{figure}

The dependence of the total cross section on the center-of-mass energy~$\sqrt{s}$ is plotted in Fig.~\ref{fig2}. One can see that its values are comparable to those of the process $\nu_l\gamma\rightarrow\nu_l\gamma\gamma$~\cite{abada2} at $\sqrt{s}\approx m_{\pi}$ exceeding the latter by about two orders of magnitude at higher energies.

Conclusions regarding the significance of the reaction (\ref{reac1}) for particle physics are the following. Information on the formfactors $F_V$ and $F_A$ is important for testing the standard model, their theoretical values depend on a model and vary in a relatively wide region \cite{Chen2011}. The reaction  (\ref{reac1}) allows to experimentally measure the formfactors in a wider range of the values of $t$ than it is possible in the decay  $\pi_{e2\gamma}$. For example, (\ref{reac1}) can be studied in scattering of neutrinos from nuclei exploiting the equivalent photon flux of the latter. In this connection another interesting problem concerning the universality of the equivalent photon approximation as a particular case of the parton model arises \cite{mine1}.

\subsection{The Primakoff effect in weak interactions}

Let us consider the reaction (\ref{reac2}), which is in fact the Primakoff effect~\cite{primakoff}  proceeding due to weak interactions. 

It is convenient to invoke the Conserved Vector Current hypothesis to find the amplitude for (\ref{reac2}), which then can be easily obtained from (\ref{ampl_decay}) by setting the formfactor $F_A=0$ (and, of course, simultaneously removing $V_{ud}$ and replacing the final state electron with a neutrino):

\begin{equation}
M=\frac{eG_{F}}{\sqrt{2}m_{\pi}}F_{V}\varepsilon_{\mu }
\bar{u}(p'_{\nu})\gamma _{\alpha }(1-\gamma
_{5})u(p_{\nu})
\epsilon ^{\mu \alpha \beta
\lambda }q_{\beta }p_{\pi \lambda }. \label{ampl_2}
\end{equation}   

One can see that (\ref{ampl_2}) is in agreement with the Lagrangian from \cite{Harvey2007} where processes related to (\ref{reac2}) by crossing have been studied.

It is now straightforward to find the cross sections of (\ref{reac2}) and its crossed reactions using the results of the Subsec.~\ref{subsec:charged}. 
Thus, setting $F_A=0$ in (\ref{cross_sect}) one obtains the differential cross section of (\ref{reac2}):

\begin{equation}
\frac{d\sigma}{dt}=-\frac{\alpha G^2_{F}}{8m^2_{\pi}}|F_V|^2\,t\left(1+\frac{u^2}{s^2}\right) \label{cross_sect2}
\end{equation}

which leads, after the integration over $t$, to the total cross section 

\begin{equation}
\sigma=\frac{\alpha G^2_{F}}{96m^2_{\pi}}|F_{V}|^2\,s^2\left(1-\frac{m^2_{\pi}}{s}\right)^2\left(\left(1-\frac{m^2_{\pi}}{s}\right)^2+6\right). \label{cross_sect_tot2}
\end{equation}

The dependence of the total cross section on $\sqrt{s}$ is plotted in Fig.~\ref{fig2}. Conclusions regarding the role playing by the reaction (\ref{reac2}) for evolution of a supernova as well as for particle physics is the same as in the Subsec.~\ref{subsec:charged}. But there is also a notable distinctive feature owing to the dominant decay of the final state neutral pion into the $\gamma\gamma$ pair. If the reaction (\ref{reac2}) proceeds in the interiors of the star than it will be accompanied by radiation of photons with change in their spectrum at the energy $m_{\pi}/2\approx67.5$~MeV.

\subsection{Annihilation of the $\bar \nu_ee^-$ pair into a charged pion}

It is also interesting to consider some of the crossed reactions of (\ref{reac1}) and (\ref{reac2}) like the following annihilation process:

\begin{equation}
\bar\nu_ee^-\rightarrow\gamma\pi^-, \label{annihil}
\end{equation}

One can similarly calculate the total cross section of the reaction (\ref{annihil}) as above:

\begin{equation}
\sigma=\frac{\alpha G^2_{F}}{24m^2_{\pi}}|V_{ud}|^2\left(|F_V|^2+|F_A|^2\right)\,s^2\left(1-\frac{m^2_{\pi}}{s}\right)^3. \label{cross_s_tann}
\end{equation}

The dependence of the cross section on $\sqrt{s}$ is shown in Fig.~\ref{fig3}.

As can be seen from (\ref{cross_s_tann}), an important advantage of using the process (\ref{annihil}) is that it determines the sum of squares of the formfactors:

\begin{equation}
\frac{24m^2_{\pi}}{\alpha G^2_{F}|V_{ud}|^2}s^{-2}\left(1-\frac{m^2_{\pi}}{s}\right)^{-3}\sigma=|F_V|^2+|F_A|^2. \label{cross_s_tann1}
\end{equation}

\subsection{Annihilation of the $\nu_l\bar \nu_l$ pair into a neutral pion}

Let us also consider the following annihilation process:

\begin{equation}
\nu_l\bar\nu_l\rightarrow\gamma\pi^0. \label{annihil2}
\end{equation}

\begin{figure}
\centering
\resizebox{0.9\textwidth}{!}{%
\includegraphics{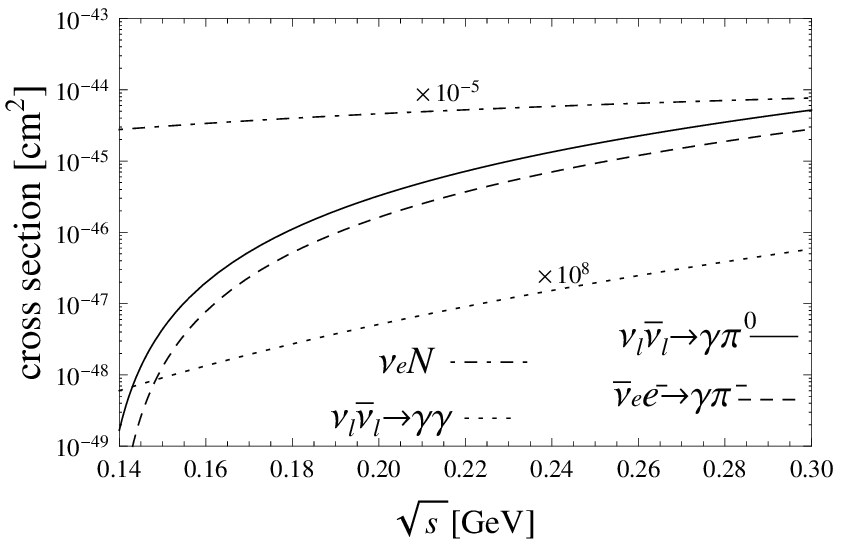}
}\caption{}
\label{fig3}
\end{figure}

Its total cross section calculated exactly as in the case of the reaction (\ref{annihil}) reads

\begin{equation}
\sigma=\frac{\alpha G^2_{F}}{12m^2_{\pi}}|F_V|^2\,s^2\left(1-\frac{m^2_{\pi}}{s}\right)^3. \label{cross_s_tann2}
\end{equation}

and the corresponding dependence on $\sqrt{s}$ is plotted in Fig.~\ref{fig3}.

In addition to similar processes such as $\nu_l\bar\nu_l\rightarrow\gamma\gamma$ \cite{abbas}, the process~(\ref{annihil2}) is also of some interest for cosmology since it can in principle unveil the cosmic neutrino background remaining still experimentally undetected. At the same time there are at least two advantages of the process (13) over $\nu_l\bar\nu_l\rightarrow\gamma\gamma$. The first one is the magnitude of its cross section exceeding the corresponding quantity of $\nu_l\bar\nu_l\rightarrow\gamma\gamma$ by factor $10^8-10^{10}$ in the considered energy range (see Fig.~\ref{fig3}). The second one is the fact that the identification of the final state photons appearing through  excitation of the $\pi^0$ mesons and their subsequent decays is more unambiguous due to the well known distinctive features of the $\gamma\gamma$ invariant mass spectrum expected in this case.

\section{Conclusions}

The exclusive reactions $\nu_e\gamma\rightarrow e^-\pi^+$, $\nu_l\gamma\rightarrow\nu_l\pi^0$,  $\bar\nu_ee^-\rightarrow\gamma\pi^-$, $\nu_l\bar\nu_l\rightarrow~\gamma\pi^0$ have been studied within the standard model. Neglecting the helicity suppressed terms $\sim O(m^2_{e,\nu}/m^2_{\pi})$, the corresponding cross sections are calculated analytically in the Fermi approximation. Some potential astrophysical implications are discussed. For example, it is shown that the pions can be produced in the interiors of a supernova in the reactions $\nu_e\gamma\rightarrow e^-\pi^+$,  $\nu_l\gamma\rightarrow\nu_l\pi^0$ whose role for evolution of the star being less important than that of neutrino--nucleon scattering is however comparable to the corresponding significance of production of structureless particles in $\nu\gamma$ interactions. The reaction $\nu_l\gamma\rightarrow\nu_l\pi^0$ can be identified by change in the spectrum of photons at about 67.5~MeV appearing due to the decay of the final state neutral pions into the $\gamma\gamma$ pairs.

It is also emphasized that $\nu_e\gamma\rightarrow e^-\pi^+$, $\bar\nu_ee^-\rightarrow\gamma\pi^-$ are crossed reactions of the decay $\pi ^{+} \rightarrow e^{+}\nu_{e}\gamma$ and therefore  can also be used to test the $V-A$ structure of the weak interaction as well as to search for physics beyond the standard model covering however a wider kinematical region than it is possible in the case of the mentioned charged pion decay.  
Though only neutrino--photon reactions have been considered, all the calculations and discussion of this paper will be obviously exactly the same for the corresponding antineutrino--photon cases provided CP is conserved.
Moreover, other pseudoscalar mesons can be treated in similar way. In particular, obtaining the cross section for the reaction $\nu_e\gamma\rightarrow e^-K^+$ is straightforward. It is enough to replace the formfactors and the mass of $\pi^+$ in (\ref{cross_sect_tot}) by the corresponding quantities of $K^+$ and to insert the CKM matrix element $V_{us}$ instead of $V_{ud}$.
Likewise, production of the neutral kaon $K^0$ can be considered, the latter may be interesting due to its participation in CP violating interactions. 
The formalism of this paper is also applicable to production of hypothetical pseudoscalar particles like axions. 

\vskip 1cm

{\bf Acknowledgements}

I thank F. F. Karpeshin for useful discussions. This work was supported in part by the Russian Foundation for Basic Research (grant 11-02-12043), by the Program for Basic Research of the Presidium of the Russian Academy of Sciences "Neutrino Physics and Neutrino Astrophysics" and by the Federal Target Program  of the Ministry of Education and Science of Russian Federation "Research and Development in Top Priority Spheres of Russian Scientific and Technological Complex for 2007-2013" (contract no.
16.518.11.7072).

\newpage

{\bf Figure Captions}
\vskip 0.5 cm 

{\bf Figure 1:} Feynman diagrams contributing to the reaction $\nu_e\gamma\rightarrow~e^-\pi^+$. The diagrams (a) and (b) are helicity suppressed, (c) depends on the structure of the $\pi$ meson.

\vskip 0.5 cm

{\bf Figure 2:} Dependence of the total cross sections of the reactions $\nu_l\gamma\rightarrow~\nu_l\pi^0$ and $\nu_e\gamma\rightarrow e^-\pi^+$ on the center-of-mass energy $\sqrt{s}$. The cross sections of the reaction $\nu_l\gamma\rightarrow\nu_l\gamma\gamma$~\cite{abada2} and quasielastic neutrino--nucleon scattering~($\nu_eN$)~\cite{neutr_nucl} (multiplied by $10^{-5}$) are shown for comparison.

\vskip 0.5 cm

{\bf Figure 3:} Dependence of the total cross sections of the reactions $\nu_l\bar\nu_l\rightarrow~\gamma\pi^0$ and $\bar\nu_ee^-\rightarrow\gamma\pi^-$ on the center-of-mass energy $\sqrt{s}$. The cross sections of the reaction $\nu_l\bar \nu_l\rightarrow\gamma\gamma$~\cite{abbas} (multiplied by $10^{8}$) and quasielastic neutrino--nucleon scattering~($\nu_eN$)~\cite{neutr_nucl} (multiplied by $10^{-5}$) are shown for comparison.


\newpage

\end{document}